# A Topological Design Tool for the Synthesis of Antenna Radiation Patterns

M. Barbuto, M.-A. Miri, A. Alu, F. Bilotti, and A. Toscano

*Abstract*—**Patch antennas are among the most popular radiating elements, yet their quasi-2D structure reduces the degrees of freedom available to tailor their radiation pattern. To overcome this limitation, a possible solution consists in etching on a grounded substrate two concentric radiating elements and combining two modes (one for each element) with proper amplitude/phase relations. Although this technique leads, in principle, to an infinite number of possible configurations (i.e., each patch element can support an infinite number of modes), the theoretical and experimental verifications available in the literature are limited to the first two radiating modes ($TM_{11}$ and $TM_{21}$) of a circular patch antenna. Recently, we have shown that the design of this circular patch can be effectively performed by exploiting the topological properties of vortex fields and, in particular, by controlling the phase singularity exhibited by the higher order right-handed circularly polarized (RHCP) $TM_{21}$ mode of the circular patch. Since the number of RHCP higher-order modes of a circular patch is infinite, we can in principle deal with an arbitrary number of phase singularity points, whose combined control leads to unprecedented possibilities to shape the radiation pattern of a circular patch. In this paper, we present a complete design tool to determine the number and position of phase singularity points arising when combining the RHCP modes of a circular patch antenna and, eventually, manipulate them to synthesize the required radiation pattern. As a realistic application example, we show how the proposed tool can be used to effectively design a single antenna whose radiation pattern can be properly tailored to switch between two different states, i.e. a sector and a saddle shape, widely used in base stations for mobile and satellite communications, respectively.**

## I. INTRODUCTION

IN several communication systems, the radiation pattern of the radiating element has to be shaped such that most of the radiated power covers a desired angular region. Common examples are base stations for mobile communications and satellite communications systems. In the first case, the radiation pattern has a sectorial shape with a wide azimuthal and a narrower elevation beamwidth. This allows reducing interference from/to other sources and focusing power on the users' plane [1]. In the second case, the antennas typically have a saddle-shaped pattern in order to compensate for the different free space loss between the satellite and the user terminals and, thus, equally distributing power over a desired portion of the Earth surface [2]. In the literature, there are several solutions to control the radiation pattern of a radiating structure based, for instance, on planar antenna arrays [3], parasitic elements [4]-[5], shaped reflectors [6]-[7], frequency selective and high impedance surfaces [8]-[9], metamaterials [10] or modulated metasurfaces [11]. However, they typically require complex feeding networks or have a limited reconfigurability. Moreover, only few of them concern single patch antennas and they typically provide lesser flexibility. In particular, a possible solution to manipulate the radiation pattern of patch antennas consists of integrating, on a single dielectric substrate, two metallic patches working in two different radiating modes [12]. However, in this way we can only switch between two different states corresponding to individually excite one or the other element. On the contrary, a more flexible solution consists on combining different radiating modes, which are simultaneously excited on a single radiating structure [13]-[14]. In this case, the amplitude of the two modes can be properly tuned to change the shape of the overall radiation pattern, while their relative phase can be exploited to choose its peak orientation.

Recently, the possibilities offered by this solution have been explained in terms of topological properties of vortex fields [15]-[16]. In particular, we have shown that the position of the single phase singularity point in the field radiated by a right-handed circular polarized (RHCP) $TM_{21}$ mode of a patch antenna can be controlled by acting on the amplitude and phase of a superimposed vortex-free RHCP $TM_{11}$ mode. This has also a direct impact on the radiation pattern of the overall structure that, thus, can be manipulated by properly acting on the topological properties of the vortex mode. Although this solution is a simple and effective technique, it has been only partially investigated. In fact, patch antennas support an infinite number of higher order modes but this solution has been tested only in the simplest case of the superposition between the fundamental mode and the first higher order one, in which only

Manuscript received Month Day, 2019.
M. Barbuto is with the "Niccolò Cusano" University, 00166 Rome, Italy (email: mirko.barbuto@unicusano.it).
M.-A. Ali and A. Alù are with the City University of New York, New York, USA.
F. Bilotti and A. Toscano are with the Department of Engineering, "Roma Tre" University, 00146 Rome, Italy.

This work has been developed in the frame of the activities of the research contract CYBER-PHYSICAL ELECTROMAGNETIC VISION: Context-Aware Electromagnetic Sensing and Smart Reaction, funded by the *Italian Ministry of Education, University and Research* as a PRIN 2017 project (protocol number 2017HZJXSZ).

one phase singularity point can be exploited.

The main contribution of this paper is to further explore the possibilities offered by this technique by studying the more general case of a circular patch antenna that simultaneously supports any two different modes, both working under RHCP. In particular, an analytical study will be carried out for determining the number and the position of the phase singularities in the overall radiated field. Moreover, since each phase singularity is inherently related to an amplitude null, we will show that this information makes it possible to properly engineer the radiating structure to obtain a desired radiation pattern with a given number and position of peaks and nulls. In other words, under certain conditions and for a given application target, we will be able to replace an antenna array by a single properly engineered radiating element.

It is interesting to note that, while the resulting design is nothing more than the combination of RHCP modes of a circular patch antenna, the result itself would not have been so easily and straightforwardly achieved without the proposed topological interpretation of the radiated field distributions and the related design tool presented in the paper. As a notable example, we focus on the simultaneous excitation of $TM_{11}$ and $TM_{31}$ RHCP modes of a circular patch antenna to show that, without increasing the complexity of the feeding network, the flexibility of the overall radiating system can be increased such that the antenna pattern can be switched from a sectorial- to a saddle-shaped one, and vice versa.

The paper is organized as follows. Section II is devoted to present the general mathematical framework for the analysis of the generic configuration. In Section III, we provide the analytical results for the specific case of the superposition of a RHCP $TM_{11}$ mode and a RHCP $TM_{31}$ mode. In Section IV, our study is validated through a proper set of full-wave numerical simulations. Finally, Section V is devoted to the conclusions.

## II. ANALYTICAL STUDY

As shown in [13]-[14], two concentric radiating patches can be used to design a simple array exhibiting radiation pattern reconfigurability and, at the same time, minimize the complexity of the feeding network. This result has been related to the topological property of a phase singularity point, which cannot be suppressed by a constant background, but only moved to another point of zero intensity. Moreover, as the shape of the radiation pattern is related to the phase singularity points of the radiated field [15], we expect that by increasing the number of phase singularities, we can also increase the degrees of freedom for shaping the radiation pattern of the overall structure.

A possible solution for generating different phase singularity points is to superimpose different vortex modes with, at least one of them, a phase singularity of order greater than one [17]. In the case of a circular patch antenna, RHCP $TM_{nm}$ modes exhibit a phase singularity of order $n-1$ [18] and, thus, a radiating mode whose order is higher than two can be used. Therefore, we need to expand the study reported in [15], which is limited to the superposition of a RHCP $TM_{11}$ mode, i.e. a constant background, and a RHCP $TM_{21}$ mode, i.e. a vortex mode with a phase singularity of the first order. For this reason, in this Section, we present the mathematical formulation for deriving the total far-field generated by the superposition of two generic RHCP $TM_{nm}$ modes radiated by a patch antenna in order to determine the number and the positions of the phase singularity points in the overall radiated field.

We start our analysis by considering the electric field radiated by a $TM_{nm}$ mode of a circular patch antenna [19]:

$$E_{\theta n} = j^n \frac{Vk_0a}{2} \frac{e^{-jk_0r}}{r} \cos n\varphi \big[ J_{n+1}(\gamma) - J_{n-1}(\gamma) \big]$$
$$E_{\varphi n} = j^n \frac{Vk_0a}{2} \frac{e^{-jk_0r}}{r} \cos\theta \sin n\varphi \big[ J_{n+1}(\gamma) + J_{n-1}(\gamma) \big]$$

(1)

where $V = hE_0 J_n(ka)$ is the edge voltage at $\varphi = 0$, $h$ is the thickness of the dielectric substrate, $E_0$ is the value of the electric field at the edge of the patch, $\gamma = k_0a\sin\theta$, $a$ is the radius of the patch, $J_i$ is the Bessel function of the first kind and order $i$, and $r$, $\theta$, and $\varphi$ are the polar coordinates.

As demonstrated in [18], a $TM_{nm}$ mode radiated by a circular patch antenna exhibits a phase singularity if excited with a circular polarization. Thus, we generalize this expression considering that a RHCP $TM_{nm}$ mode can be expressed as a superposition of the individual electric fields produced by two linearly polarized orthogonal modes with a 90° phase shift between them:

$$E_{\theta n}^t = E_{\theta n}^1(\phi,\theta) + jE_{\theta n}^2(\phi + \tau,\theta)$$
$$E_{\varphi n}^t = E_{\varphi n}^1(\phi,\theta) + jE_{\varphi n}^2(\phi + \tau,\theta)$$

(2)

where superscripts 1 and 2 correspond to the fields generated by the two orthogonal modes and $\tau$ is the angular spacing of the feeds, which depends on the mode order, as reported in [20].

Moreover, since we are interested in the superposition of two generic RHCP $TM_{nm}$ modes (indicated with azimuthal indices $n_1$ and $n_2$, with $n_1 < n_2$), we calculate the overall radiated field as a weighted sum of these two fields expressed according to (2):

$$E_\theta^t = \sin(\alpha)\big[ E_{\theta n_1}^1(\phi,\theta) + jE_{\theta n_1}^2(\phi + \tau_1,\theta) \big] +$$
$$+ \cos(\alpha)\big[ E_{\theta n_2}^1(\phi,\theta) + jE_{\theta n_2}^2(\phi + \tau_2,\theta) \big] e^{j\delta}$$
$$E_\varphi^t = \sin(\alpha)\big[ E_{\varphi n_1}^1(\phi,\theta) + jE_{\varphi n_1}^2(\phi + \tau_1,\theta) \big] +$$
$$+ \cos(\alpha)\big[ E_{\varphi n_2}^1(\phi,\theta) + jE_{\varphi n_2}^2(\phi + \tau_2,\theta) \big] e^{j\delta}$$

(3)

where the terms in square brackets are the field components related to the two modes, and $\alpha$ and $\delta$ specify their relative amplitude and phase, respectively.

As shown in [18], in Cartesian coordinates relations, equations (1) for circular polarization operation can be simplified to:

$$E_{xn} = A_n e^{-j(n-1)\phi} - B_n e^{-j(n+1)\phi}$$
$$jE_{yn} = A_n e^{-j(n-1)\phi} + B_n e^{-j(n+1)\phi} \tag{4}$$

with amplitudes:

$$A_n = -j^n \frac{V_n k_0 a_n}{2} \frac{e^{-jk_0 r}}{r} \cos\theta J_{n-1}\left(\gamma_n\right)$$
$$B_n = -j^n \frac{V_n k_0 a_n}{2} \frac{e^{-jk_0 r}}{r} \cos\theta J_{n+1}\left(\gamma_n\right) \tag{5}$$

Therefore, the total field (3) for generic order modes reduce to:

$$E_x = \sin\alpha\left(A_{n_1} e^{-j(n_1-1)\phi} - B_{n_1} e^{-j(n_1+1)\phi}\right) + e^{j\delta}\cos\alpha\left(A_{n_2} e^{-j(n_2-1)\phi} - B_{n_2} e^{-j(n_2+1)\phi}\right)$$
$$jE_y = \sin\alpha\left(A_{n_1} e^{-j(n_1-1)\phi} + jB_{n_1} e^{-j(n_1+1)\phi}\right) + e^{j\delta}\cos\alpha\left(A_{n_2} e^{-j(n_2-1)\phi} + jB_{n_2} e^{-j(n_2+1)\phi}\right) \tag{6}$$

Using (5) in (6), we get:

$$\left(\frac{k_0}{2}\frac{e^{-jk_0 r}}{r}\right)^{-1} E_x = -j^{n_1} V_{n_1} a_{n_1} \sin\alpha\left(J_{n_1-1}\left(\gamma_{n_1}\right)e^{-j(n_1-1)\phi} - J_{n_1+1}\left(\gamma_{n_1}\right)e^{-j(n_1+1)\phi}\right)\cos\theta +$$
$$-j^{n_2} e^{j\delta} V_{n_2} a_{n_2} \cos\alpha\left(J_{n_2-1}\left(\gamma_{n_2}\right)e^{-j(n_2-1)\phi} - J_{n_2+1}\left(\gamma_{n_2}\right)e^{-j(n_2+1)\phi}\right)\cos\theta$$
$$j\left(\frac{k_0}{2}\frac{e^{-jk_0 r}}{r}\right)^{-1} E_y = -j^{n_1} V_{n_1} a_{n_1} \sin\alpha\left(J_{n_1-1}\left(\gamma_{n_1}\right)e^{-j(n_1-1)\phi} + jJ_{n_1+1}\left(\gamma_{n_1}\right)e^{-j(n_1+1)\phi}\right)\cos\theta +$$
$$-j^{n_2} e^{j\delta} V_{n_2} a_{n_2} \cos\alpha\left(J_{n_2-1}\left(\gamma_{n_2}\right)e^{-j(n_2-1)\phi} + jJ_{n_2+1}\left(\gamma_{n_2}\right)e^{-j(n_2+1)\phi}\right)\cos\theta \tag{7}$$

Limiting our attention to small deviations from the propagation axis $z$, we can consider $\gamma_n = k_0 a_n \sin\theta \ll 1$. Therefore, in each bracket, one of the Bessel functions can be neglected if compared to the other one, i.e., $J_{n_1-1}\left(\gamma_{n_1}\right) \gg J_{n_1+1}\left(\gamma_{n_1}\right)$ and $J_{n_2-1}\left(\gamma_{n_2}\right) \gg J_{n_2+1}\left(\gamma_{n_2}\right)$, which, in turns, results in the simplified relations:

$$\left(\frac{k_0}{2}\frac{e^{-jk_0 r}}{r}\right)^{-1} E_x = -j^{n_1} V_{n_1} a_{n_1} \sin\alpha\left(J_{n_1-1}\left(\gamma_{n_1}\right)e^{-j(n_1-1)\phi}\right)\cos\theta - j^{n_2} e^{j\delta} V_{n_2} a_{n_2}\cos\alpha\left(J_{n_2-1}\left(\gamma_{n_2}\right)e^{-j(n_2-1)\phi}\right)\cos\theta$$
$$j\left(\frac{k_0}{2}\frac{e^{-jk_0 r}}{r}\right)^{-1} E_y = -j^{n_1} V_{n_1} a_{n_1} \sin\alpha\left(J_{n_1-1}\left(\gamma_{n_1}\right)e^{-j(n_1-1)\phi}\right)\cos\theta - j^{n_2} e^{j\delta} V_{n_2} a_{n_2}\cos\alpha\left(J_{n_2-1}\left(\gamma_{n_2}\right)e^{-j(n_2-1)\phi}\right)\cos\theta \tag{8}$$

Therefore, under this approximation, the field components are both proportional to:

$$-j^{n_1} V_{n_1} a_{n_1}\sin\alpha\left(J_{n_1-1}\left(\gamma_{n_1}\right)e^{-j(n_1-1)\phi}\right)\cos\theta +$$
$$-j^{n_2} e^{j\delta} V_{n_2} a_{n_2}\cos\alpha\left(J_{n_2-1}\left(\gamma_{n_2}\right)e^{-j(n_2-1)\phi}\right)\cos\theta \tag{9}$$

Considering again $\gamma_n \ll 1$, one can write:

$$J_{n-1}\left(\gamma_n\right) \simeq \frac{1}{(n-1)!}\left(\frac{\gamma_n}{2}\right)^{n-1} \tag{10}$$

Therefore, (9) becomes:

$$-j^{n_1} V_{n_1} a_{n_1}\sin\alpha \frac{\left(\gamma_{n_1}\right)^{n_1-1}}{2^{n_1-1}\left(n_1-1\right)!} e^{-j(n_1-1)\phi}\cos\theta +$$
$$-j^{n_2} V_{n_2} a_{n_2}\cos\alpha \frac{\left(\gamma_{n_2}\right)^{n_2-1}}{2^{n_2-1}\left(n_2-1\right)!} e^{j\left[\delta-(n_2-1)\right]\phi}\cos\theta \tag{11}$$

and explicating $\gamma_n$ we obtain:

$$-j^{n_1} V_{n_1} a_{n_1}\sin\alpha \frac{1}{(n_1-1)!}\left(\frac{k_0 a_{n_1}\sin\theta}{2}\right)^{n_1-1} e^{-j(n_1-1)\phi}\cos\theta - j^{n_2} V_{n_2} a_{n_2}\cos\alpha \frac{1}{(n_2-1)!}\left(\frac{k_0 a_{n_2}\sin\theta}{2}\right)^{n_2-1} e^{j\left[\delta-(n_2-1)\right]\phi}\cos\theta \tag{12}$$

In order to simplify this expression, we define:

$$C_{n_1} = V_{n_1} a_{n_1} \sin\alpha \, \frac{1}{(n_1-1)!} \left(\frac{k_0 a_{n_1}}{2}\right)^{n_1-1}$$
$$C_{n_1} = V_{n_2} a_{n_2} \cos\alpha \, \frac{1}{(n_2-1)!} \left(\frac{k_0 a_{n_2}}{2}\right)^{n_2-1} \tag{13}$$

obtaining, thus:

$$-j^{n_1} C_{n_1} \sin\alpha (\sin\theta)^{n_1-1} e^{-j(n_1-1)\phi} \cos\theta +$$
$$-j^{n_2} C_{n_2} \cos\alpha (\sin\theta)^{n_2-1} e^{j\left[\delta-(n_2-1)\right]\phi} \cos\theta \tag{14}$$

Finally, in order to find the location of the vortices, we can search for positions where the field components simultaneously vanish:

$$-j^{n_1} C_{n_1} \sin\alpha (\sin\theta)^{n_1-1} e^{-j(n_1-1)\phi} \cos\theta +$$
$$-j^{n_2} C_{n_2} \cos\alpha (\sin\theta)^{n_2-1} e^{j\left[\delta-(n_2-1)\right]\phi} \cos\theta = 0 \tag{17}$$

From this equation, we can see that we have a phase singularity of order $n_1$-$1$ in the origin. In fact, for $\theta{\rightarrow}0$ the second term can be neglected compared to the first one, which has the typical phase term ($e^{-j(n_1-1)\phi}$) of a vortex beam with a topological charge of order $n_1$-$1$.

Moreover, using the relation $j^n = e^{jn\pi/2}$, (17) becomes:

$$-C_{n_1} \sin\alpha (\sin\theta)^{n_1-1} e^{-j(n_1-1)\phi+jn_1\pi/2} \cos\theta +$$
$$-C_{n_2} \cos\alpha (\sin\theta)^{n_2-1} e^{j\left[\delta-(n_2-1)\right]\phi+jn_2\pi/2} \cos\theta = 0 \tag{18}$$

and with a few mathematical manipulations, we end up with a simple expression for the other phase singularity points:

$$\phi = \frac{\delta + (n_2-n_1)\pi/2 + (2k-1)\pi}{n_2-n_1}$$
$$(\sin\theta)^{n_2-n_1} = \frac{C_{n_1}}{C_{n2}} \tan\alpha \tag{19}$$

where $k$ is any integer value.

From (19) we can infer that, in addition to the central phase singularity of order $n_1$-$1$, the overall radiated field exhibits $n_2$-$n_1$ external vortices. Moreover, on a given $z$-plane in the far-field, the distance of the vortices from the origin depends on $\alpha$ while their azimuth depends on $\delta$.

## III. A PRACTICAL EXAMPLE

As demonstrated in the previous Section, by properly choosing the order of two different superimposed RHCP TM$_{nm}$ modes, we can force the number and the topological charge of the phase singularity points in the overall radiated field. In particular, the order of the lower mode defines the order of the topological charge in the center beam. The difference between the two mode orders determines, instead, the number of the other phase singularity points, whose positions can be controlled by acting on the amplitude and relative phase excitation of the two modes. Furthermore, we remark here that any phase singularity point is inherently related to an amplitude null. Therefore, by acting on the same parameters, we can also control the presence and the position of radiation nulls on the overall radiated field.

As a particular case, this possibility has been tested in [15] in the case of a RHCP TM$_{11}$ mode superimposed to a RHCP TM$_{21}$ mode. However, the presence of only one singularity point does not allow to shape at will the overall radiation pattern. On the contrary, (19) and analysis previously reported, suggest that by involving higher order modes of patch antennas, a greater number of phase singularities can be exploited for advanced pattern shaping applications.

In order to present an example of practical interest and, thus, validate our approach, we now study the particular case of the superposition between a RHCP TM$_{11}$ mode, i.e. a constant background, and a RHCP TM$_{31}$ mode, i.e. a vortex mode with a phase singularity of the second order. In this case, the lower order mode has no phase singularities and, thus, we expect no amplitude null in the center of the overall radiated beam. Moreover, as preliminary reported in [21], for this combination of order modes (19) becomes:

$$\phi = \frac{\delta}{2} + k\pi$$
$$\sin^2\theta = \frac{C_1}{C_3} \tan\alpha \tag{20}$$

From these equations we can infer that the second order phase singularity of the TM$_{31}$ mode actually splits into two phase singularities of the first order, whose positions can be tailored by acting on $\alpha$ and $\delta$. In particular, for $\delta = 0$, the two singularities are symmetrically placed on the x-axis and by increasing $\alpha$ they move away from the center of the beam. Conversely, for a fixed value of $\alpha$, by increasing $\delta$ the vortices rotate around a fixed circle of an angle equal to $\delta/2$.

### A. Amplitude and phase patterns

In order to graphically check the analytical formulas (20) and get a pictorial representation of the previous trends when varying the excitation amplitude and phase, we have analyzed the phase and amplitude patterns of the radiated field for different values of $\alpha$ and $\delta$, as shown in Fig. 1 and Fig. 2. In particular, we start discussing the phase patterns of the x- and y-components on a plane normal to the propagation direction assuming $\delta = 0$ and for different values of $\alpha$ (Fig. 1). Please note that $\alpha$ specifies the relative amplitude of the two modes through the trigonometric functions in (3). Therefore, the two limiting cases will be $\alpha = 0$, corresponding to the presence of the TM$_{31}$ mode only, and $\alpha = \pi/2$, corresponding to the presence of the fundamental mode only. In the first case, there is a phase singularity of the second order in the beam center while, in the second case, the radiated field has no phase singularities. On the contrary, for $0<\alpha<\pi/2$, the second order phase singularity carried by the $TM_{31}$ mode splits into two phase singularities of the first order that move away from the center beam, due to the superposition with a constant background. Moreover, the last row of Fig. 1 confirms that any phase singularity can exist only in a point of zero intensity and, thus, we can also control the position of two dark cores that, by increasing $\alpha$, move away from the center of the radiating structure.

Then, we have focused our attention, without loss of generality, on the case $\alpha = \pi/4$ (i.e. the two modes are excited with equal amplitude) and analyzed the phase patterns of the x- and y-components of the radiated field for different values of $\delta$ (Fig. 2). From these pictures we can see that the two topological singularities arising from the $TM_{31}$ mode rotate around the center beam of an angle equal to a half of the phase shift $\delta$, as predicted by the analytical model. The same phenomenon occurs for the amplitude nulls (reported in the last row of Fig. 2) that rigidly rotate with the phase singularity points.

### B. Radiation Patterns

As previously discussed, the possibility to control the position of the phase singularities of composite vortices and, thus, of their amplitude nulls allows exploiting topological properties of vortex modes for shaping the radiation pattern of patch antennas. In order to show that increasing the number of phase singularities enhances also the possibilities offered by this technique, we have evaluated the overall 3D directivity patterns for different values of $\alpha$ and $\delta$. We start our analysis by evaluating the effect of $\alpha$ on the shape of the radiation pattern. As shown in Fig. 3, for $\alpha = 0$ the antenna radiates with the typical conical pattern of a patch antenna working on the higher-order $TM_{31}$ mode. By progressively increasing the amplitude of the fundamental mode, the centered radiation null splits into two nulls on the x-axis that move away from the center beam. In particular, by properly choosing the excitation amplitudes, the two nulls can be placed to obtain a saddle- or a sector-shaped pattern. The other limiting case, instead, is $\alpha = \pi/2$, corresponding to a broadside radiation pattern, as expected for a patch antenna working on its fundamental mode.

Then, we have focused, without loss of generality, on the case of a sector element ($\alpha = 3\pi/8$) and from the corresponding results reported in Fig. 4, it is evident that, by adding a phase shift $\delta$ between the two exciting modes, the radiation pattern rotates around the vertical axis of an amount $\delta/2$. We remark here the high flexibility in the radiation pattern shaping enabled by the proper control of the two phase singularity points. In fact, excluding the two limiting cases that correspond to standard radiation patterns of patch antennas, the three intermediate cases reported in Fig. 3 confirm that we can switch between very different radiation patterns such as a dual beam shape, the one of a sectorial element or a saddle-shaped pattern. In particular, these new configurations can find application in base stations for mobile communications and in satellite communications, where reduced costs and size are highly desirable. Moreover, we can also design a sectorial or a saddle element whose radiation pattern can be electronically rotated by simply acting on the excitation phase of the two modes.

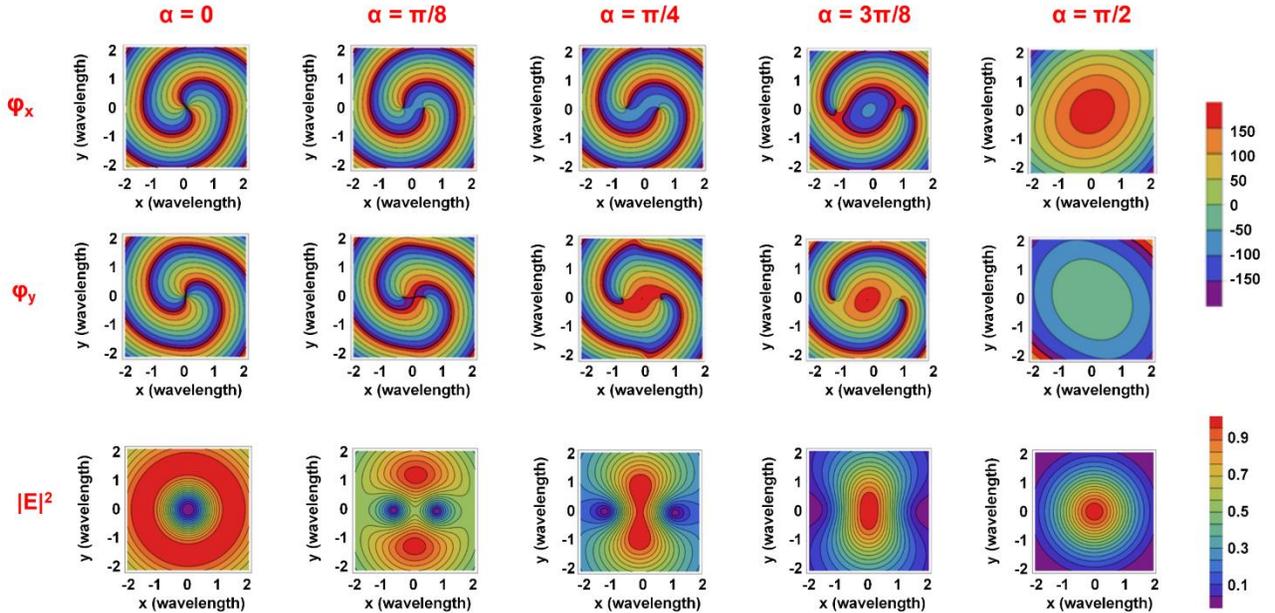

Fig. 1. Analytically calculated phase ($\Phi_x$ and $\Phi_y$, in degree) and total electric energy density ($|E|^2$) distributions at a distance $\lambda_0$ from the patch of the electric field radiated by the superposition of RHCP $TM_{11}$ and RHCP $TM_{31}$ modes for different values of $\alpha$ (here $\delta = 0$).

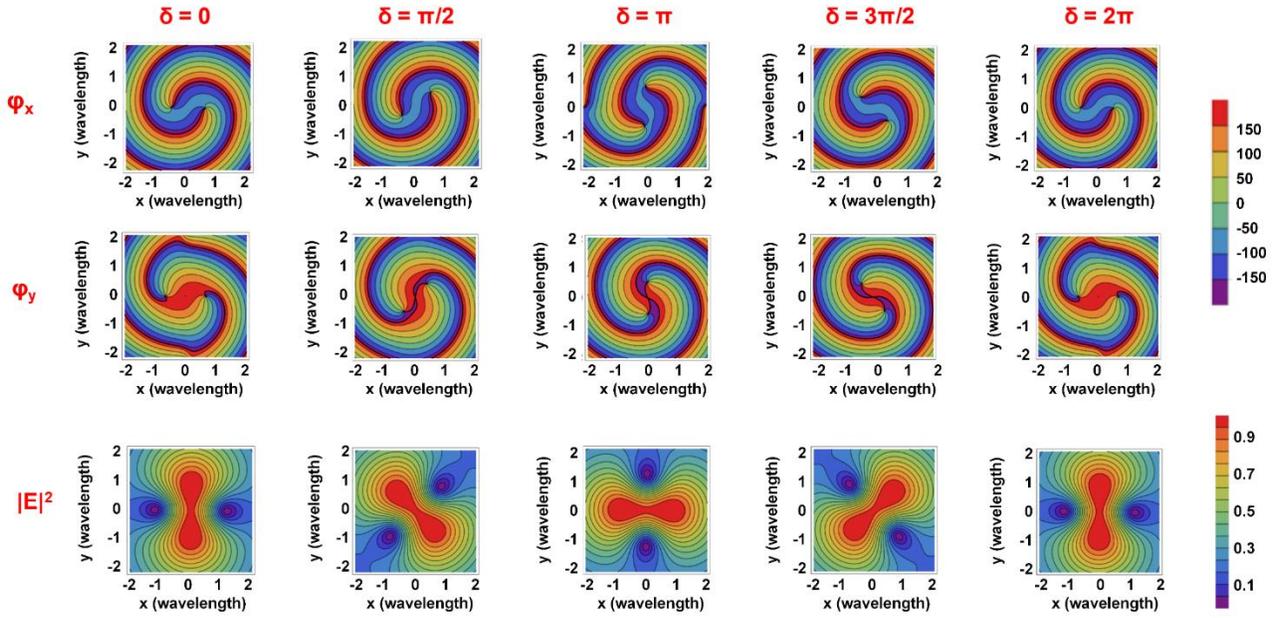

Fig. 2. Analytically calculated phase ($\Phi_x$ and $\Phi_y$, in degree) and total electric energy density ($|E|^2$) distributions at a distance $\lambda_0$ from the patch of the electric field radiated by the superposition of RHCP $TM_{11}$ and RHCP $TM_{31}$ modes for different values of $\delta$ (here $\alpha = \pi/4$).

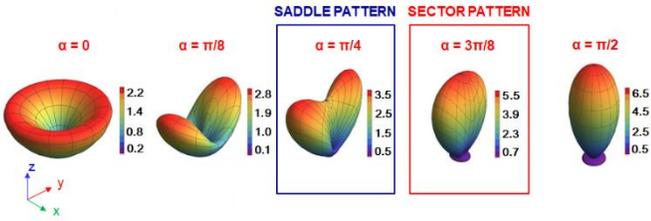

Fig. 3. Analytically calculated directivity patterns for the superposition of RHCP $TM_{11}$ and RHCP $TM_{31}$ modes for different values of $\alpha$ (here $\delta = 0$).

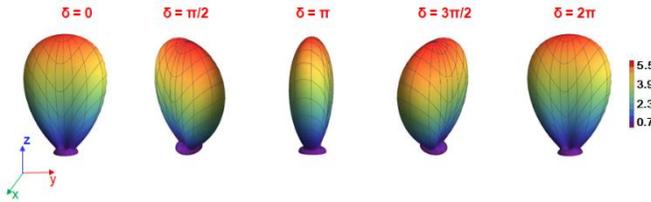

Fig. 4. Analytically calculated directivity patterns for the superposition of RHCP $TM_{11}$ and RHCP $TM_{31}$ modes for different values of $\delta$ (here $\alpha = \pi/4$).

## IV. FULL-WAVE NUMERICAL VALIDATION

As discussed in [15], in order to generate composite vortices at microwave frequencies, we need to design a single radiating element able to superimpose two different radiating modes. An effective and simple solution is to concentrically place, on a common grounded substrate, an inner circular patch and an outer annular ring. In this way, we can choose the dimensions of the two radiating elements in order to radiate the required different modes that, in the case here investigated, are a RHCP $TM_{11}$ mode and a RHCP $TM_{31}$ mode. The proposed structure is shown in Fig. 5, where, for each radiating element, we can also see two peripheral slits and a single coaxial feed properly co-designed to excite circular polarized modes around 3 GHz. In Figs. 6 and 7, the magnitude of the scattering parameters and the axial ratio for the two feeds are reported, confirming the good impedance matching and polarization purity of the two radiating elements. Then, we have analyzed the performance of the overall structure when the two elements are simultaneously excited but with different amplitudes and phases. The results for the amplitude and phase patterns, not reported here for the sake of brevity, are almost superimposed to the ones analytically predicted (Fig. 1 – Fig. 2). Moreover, the radiation patterns, reported in Fig. 8 and Fig. 9, confirm the reconfigurable capabilities provided by the topological properties of the higher order mode. In particular, we have reported here the most interesting configurations corresponding to a sectorial- and a saddle-shaped pattern, confirming the effectiveness of the proposed technique in electronically controlling the radiation pattern of a simple printed structure. Please, note that the analytical patterns reported in Fig. 3-4 have been calculated by assuming the presence of an infinite ground plane and the perfect collinear superposition of the required vortex modes. In contrast, the numerically simulated patterns reported in this Section have been obtained by taking all the constrains for realizing a realistic structure into account, i.e. a ground plane with finite dimensions and two concentrically (but not superimposed) radiating elements. This unavoidably leads to some discrepancies between analytical and numerical results that, however, do not affect the main shape characteristics of the radiation patterns.

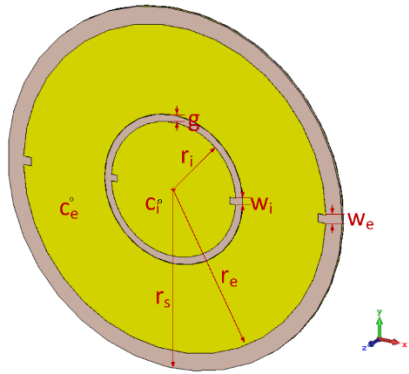

Fig. 5. Perspective view of the proposed radiating structure. Dimensions: g = 2.1 mm, $r_e$ = 45 mm, $r_i$ = 18.5 mm, $r_s$ = 50 mm, $w_e$ = 2.4 mm, $w_i$ = 1.83 mm, $c_e$ = (-30.6 mm; -8.3 mm), $c_i$ = (-4 mm; -4 mm). The radiating patches are etched on a Roger Duroid™ RT5870, with $\varepsilon_r$ = 2.33, tan δ = 0.0012, and thickness 0.787 mm.

Finally, we remark here that the investigated structure is only one of the possible configurations that can be envisioned exploiting (19). In fact, by properly selecting the radiating modes of the circular polarized radiating elements and their relative amplitude and phase of excitation, we can further manipulate the number and positions of the radiating nulls. Moreover, these new degrees of freedom can be exploited to properly shape the radiation pattern and, at the same time, reduce the complexity of the overall system compared to a conventional antenna array.

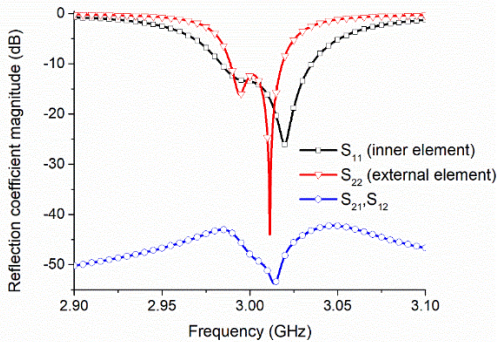

Fig. 6. Simulated magnitude of the scattering parameters at the input ports the radiating elements shown in Fig. 5.

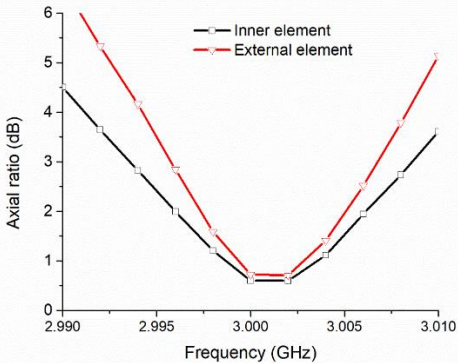

Fig. 7. Simulated axial ratios in the main beam direction of the radiating elements shown in Fig. 5.

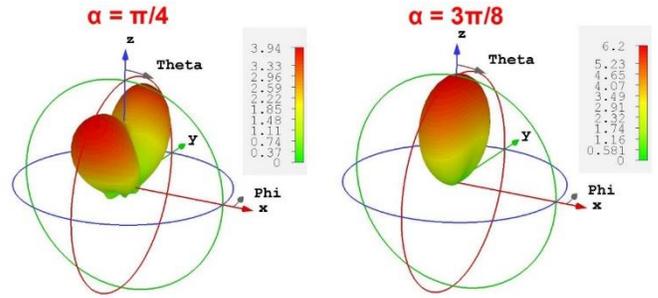

Fig. 8. Numerically calculated directivity patterns for the superposition of RHCP $TM_{11}$ and RHCP $TM_{31}$ modes for α = π/4 (left) and α = 3π/8 (here δ = 0).

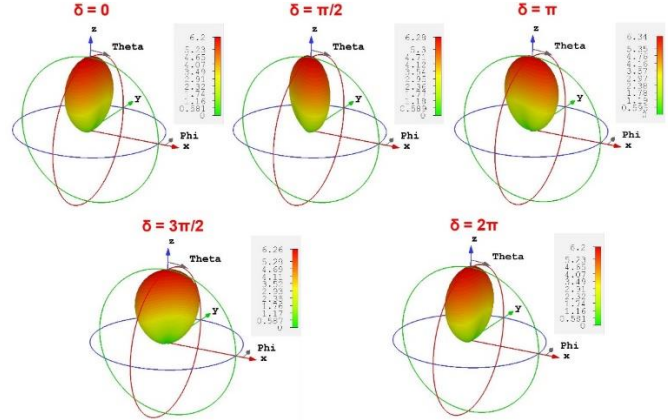

Fig. 9. Numerically calculated directivity patterns for the superposition of RHCP $TM_{11}$ and RHCP $TM_{31}$ modes for different values of δ (here α = π/4).

## V. CONCLUSION

The pattern shaping capabilities offered by composite vortices at microwave frequencies have been recently introduced for the particular case of the superposition between the first two radiating modes of a circular patch antenna. In this paper, we have extended the possibilities of this technique by considering the more general case of the superposition between generic order modes. In particular, through an analytical approach for the calculation of the overall far-field, we have shown that we can increase the number of the phase singularities exhibited by the radiated field and, thus, the degrees of freedom for an advanced and unprecedented manipulation of the overall radiation pattern. In order to validate our design approach, we have presented a practical example of an electronically rotating sectorial- or saddle-shaped pattern, which could be useful in several application fields, such as mobile communication systems or satellite radio links. The presented analytical results agree well with the full-wave simulation results and confirm that topological properties of composite vortices can be exploited as a new design strategy for shaping the radiation pattern of patch antennas.